# 3D-printed Terahertz Photonic Bandgap Waveguide-based Fluidic Sensor

YANG CAO,[1] KATHIRVEL NALLAPPAN,[1,2] HICHEM GUERBOUKHA,[1] THOMAS GERVAIS,[1] AND MAKSIM SKOROBOGATIY[1,*]

[1]*Engineering Physics, Polytechnique de Montréal, C. P. 6079, succ. Centre-ville, Montréal, Quebec H3C 3A7, Canada*
[2]*Electrical Engineering, Polytechnique de Montréal, C. P. 6079, succ. Centre-ville, Montréal, Quebec H3C 3A7, Canada*
*[\*maksim.skorobogatiy@polymtl.ca](mailto:maksim.skorobogatiy@polymtl.ca)*

**Abstract:** A Bragg waveguide-based resonant fluidic sensor operating in THz band is studied. A fused deposition modeling 3D printing technique is employed to fabricate the sensor where the liquid analyte is flowing in the microfluidic channel integrated into the waveguide cladding. The analyte refractive index-dependent resonant defect state supported by the fluidic channel is probed by tracking the resulting absorption dip and phase change of the core-guided mode on waveguide transmission spectra. The proposed fluidic sensor can open new opportunities in applied chemical and biological sensing as it offers a non-contact measurement technique for monitoring refractive index changes in flowing liquids.

## 1. Introduction

The terahertz (THz) frequency band (100 GHz – 10 THz, wavelengths 3 mm – 30 µm) holds great promise for non-destructive imaging, industrial sensing and process monitoring due to relative transparency of many important dielectric materials, low photon energy (especially important for bio-applications), as well as unique spectral fingerprints of numerical materials in this spectral range [1]. In particular, monitoring of the refractive index (RI) of fluids using THz waves has attracted much interest due to various important applications in chemistry and biology. THz fluidic sensors developed so far were used to probe the binding state of the nucleic acids [2], to monitor the conformational change and interaction of the proteins [3], as well as to detect pesticides and antibiotics [4,5].

The most sensitive detection strategy for monitoring liquid analyte RI relies on tracking the position of spectral singularities caused by the resonant excitation of a certain defect optical state by the probing THz wave [6]. Either the signal amplitude or phase can be used in such experiments, with phase being, generally, the most sensitive parameter for detection of small changes in the analyte RI. Such resonant THz fluidic sensors have been so far realized using metamaterials [7–11], reflectors with resonant cavities [12], as well as waveguide structures with or without photonic crystal reflectors [13–15].

In metamaterial-based sensors, artificially-engineered micrometer-sized inclusions are patterned on a planar substrate and subsequently brought close to liquid analytes. The electromagnetic response of the metamaterial is then probed to monitor the shift of the metamaterial resonant frequency, which can then be related to changes in the thickness and/or RI of the liquid analyte [7–11]. The best sensitivities measured to date using this methodology are ~50 GHz/µm and ~700 GHz/RIU for sensors operating at relatively high (3.5 THz and 1.5 THz) frequencies [10,11]. Due to high sensor sensitivities, the range of measured thickness and RIs are limited to ~10 µm and ~1 RIU, respectively. Unfortunately, high quality metamaterial-based sensors operating at high frequencies (~1 THz and higher) are difficult to fabricate due to significant structural complexity and processing costs, poor stability [16], and non-uniform spectral sensitivity [10,11], which are all due to the size of elemental features (such as split ring resonators) that can be as small as $\lambda/3=100$ µm at 1 THz.

For fluidic sensors based on the resonant cavity reflectors, one implementation featured a metallic cavity covered with aluminum foil onto which an array of grooves (widths 220 µm, spacing 330 µm) was inscribed. The liquid analyte was introduced into the grooves, and the reflection spectrum of such a structure showed absorption dips corresponding to the excitation of the cavity resonances. By detecting the spectral position of the absorption dips, the sensor featured sensitivity of ~500 GHz/RIU near ~0.6 THz and offered 0.4 RIU operational range; however, the practical application of this particular sensor was limited due to its fragile structure, lack of fluidic control and vicinity of its operational frequency to the water absorption peak [12].

In contrast, waveguide-based sensors can be fabricated using simpler geometries and more cost-effective techniques. Thus, in [13], a parallel-plate waveguide (PPWG) with a 1mm gap was used to probe resonantly the RI of liquid analytes filling the groove that was inscribed onto the surface of a waveguide hollow core. The groove featured rectangular cross section with width and depth of ~400 µm, and functioned as a resonant cavity for the THz wave propagating perpendicularly to it. A sensitivity of ~90 GHz/RIU was measured in the 0.27 – 0.3 THz frequency range, and offered ~0.3 RIU measurement range. The same research group then proposed a multichannel structure with similar architecture for multi-sample detection [14]. Furthermore, in [15], a hollow-core tube-shaped THz waveguide with a grating (period 300 μm) inscribed on its outer surface that also contained liquid analyte was employed as a fluidic sensor featuring sensitivity of ~50 GHz/μL at ~0.6 THz. Changes in the grating RI impacted the transmission properties of the core-guided THz wave of the ARROW waveguide. Changes in the analyte RI can then be extracted from changes in the frequency of maximal transmission efficiency through the tube waveguide.

Among all the resonant sensors of RI, hollow-core photonic bandgap (PBG) Bragg waveguides offer many practical advantages including simplicity and flexibility of design and fabrication [17], high sensitivities, ease of integration into sensing systems, integration of microfluidics directly into the waveguide structure, as well as operation range from visible-near-IR [18–20] to THz spectral range [21]. By introducing a geometrical or RI defect inside the waveguide periodic reflector, a spectral singularity in the transmitted spectra is introduced at the frequency of anticrossing between the dispersion relations of the core-guide mode and the defect state in the reflector [22,23]. Thus, integrating a fluidic channel inside of a PBG waveguide reflector allows real-time resonant monitoring of the analyte RI by tracking the position of the anticrossing frequency that normally manifests itself as a narrow dip in the waveguide transmission spectrum.

From the device fabrication and system design point of view, THz sensing components operating below ~0.5 THz are relatively easy to fabricate due to large wavelengths in the THz band, and they also profit from relatively low atmospheric absorption (no major water vapor peaks at lower frequencies) thus forgoing the need for nitrogen enclosures. It is, therefore, no surprise that 3D printing that offers 50 – 400 μm resolutions even when using low-cost tabletop systems has emerged as one of the prime tools for rapid, cost-effective fabrication of the THz components for operation at lower THz frequencies [21,24–28].

In this work, we present theoretical analysis and experimental study of a THz hollow-core Bragg waveguide-based resonant fluidic sensor. The spectral position of the waveguide reflector bandgap and anticrossing frequency of its fundamental mode with that of a defect state in the reflector are first studied numerically using finite element COMSOL Multiphysics software. Then, we explore several fabrication strategies for such sensors using fused deposition modeling (FDM) and stereolithography (SLA) techniques, and comment on their comparative advantages and disadvantages. We then experimentally optimize the sensor by choosing the proper waveguide length and the optimal position of the reflector defect, as well as comment on the limitations in the sensor resolution posed by the accuracy of additive manufacturing techniques used in sensor fabrication. Next, using a continuous-wave (CW) THz spectroscopy system, the THz spectral response of an optimized Bragg waveguide-based sensor is studied for different values of the liquid analyte (different oils) refractive indices. By tracking the position of the spectral dip caused by the anticrossing phenomenon, our sensor sensitivity is found to be ~110 GHz/RIU near 0.23 THz, which is in good agreement with the theoretical result. By analyzing errors in the spectral response curves, we also find that our sensor resolution is ~$4.5 \cdot 10^{-3}$ RIU. We believe that our THz resonant fluidic sensors can be applied in a variety of applications that require monitoring of RI changes in otherwise non-optically transparent liquid analytes.

## 2. Theoretical analysis of the THz hollow-core Bragg waveguides

*2.1 Hollow-core Bragg waveguide with a perfectly periodic reflector*

Over the past twenty years, a number of Bragg fibers were fabricated using the fiber drawing technique and offering operation in the visible [18–20,29,30], near infrared (IR) [31,32] and mid-IR [33] spectral ranges. Such fibers typically operate within the photonic bandgap of a periodic reflector in the fiber cladding by confining the guided light in the fiber hollow core. In the reflector region, Bragg fibers feature periodic multilayers with the individual layer thicknesses in the sub-μm (visible) to several μm (mid-IR) range. A typical size of the fiber hollow core is in the several hundred μm range [18–20,29–33]. Most recently, several hollow-core Bragg waveguide designs were demonstrated in the THz spectral range using mechanical rolling [34,35] and 3D printing fabrication techniques [21]. Due to much longer THz wavelengths compared to the IR wavelengths, Bragg fibers designed for THz applications are much easier to fabricate as they feature much larger core sizes and reflector layer thicknesses.

While most Bragg fibers in the visible and IR spectral ranges feature circular cross section, which is the shape most suitable for fiber drawing, in the THz spectral range this might not be the optimal shape for fabrication. This is due to the fact that THz Bragg waveguides are more convenient to fabricate using various 3D printing techniques such as FDM and SLA, that tend to produce linear structures (lines, planes and walls) of much higher quality than the curved structures (circles, spheres, etc.). This is related to the fact that most FDM printers are built with a Cartesian system in mind with mechanical parts (extruders and a built plane) that move naturally along the three orthogonal directions [36]. At the same time, most of the tabletop SLA devices are built around a projection system that defines a pattern of UV exposed resin at the built plane using a rectangular matrix or micromirrors (DLP chip) [37]. Therefore, linear pattern aligned with the principal axis of a 3D printer hardware tends to give the highest build quality. In what follows, we therefore focus on studying square-shaped Bragg waveguides with the choice of the waveguide geometry mostly motivated by the employed fabrication method.

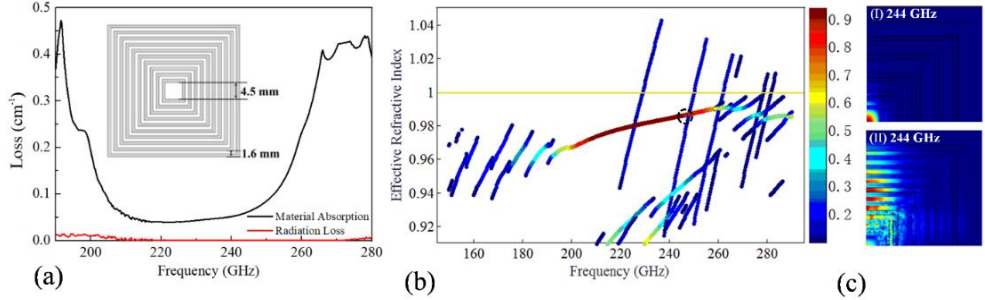

Fig. 1. (a) Spectral dependence of the material absorption and radiation loss for the fundamental core-guided mode of a hollow-core Bragg waveguide. Insert: schematic of the waveguide cross section. The grey and white layers represent the high- (plastic) and low- (air) RI materials respectively. (b) Band diagram of the guided modes in the hollow-core Bragg waveguide featuring perfectly periodic reflector. The color scale defines the fraction of the modal power confined to the waveguide hollow core. Only the modes with over 10% of power guided in the core are presented. (c) Modal field distributions of the core guided mode (I) and a Bragg reflector surface state (II) at 244 GHz near the point of their accidental degeneracy.

Particularly, the hollow-core Bragg waveguide studied in our work features a hollow core of 4.5 mm size surrounded by a periodic reflector in the cladding region consisting of twenty alternating layers of plastic and air. Each of the ten bilayers consists of a high-RI Polylactic Acid (PLA) layer and a low-RI air layer ($n_a$= 1 [21]) with equal thicknesses of 0.8 mm (see insert in Fig. 1a). First, we numerically study the modes of a defect-free Bragg waveguide using the finite element COMSOL Multiphysics software. In our simulations, we use the experimental value of the PLA material RI $n_{PLA}$=1.63 and loss $\alpha_{PLA}$=0.95 cm$^{-1}$ in the vicinity of 0.2 THz as measured using the cut-back method and a CW-THz system (see Supplementary Material).

With thus defined geometrical and materials parameters, in the 205 GHz - 255 GHz spectral range, the hollow-core Bragg waveguide features a single Gaussian-like core-guided mode (see band diagram in Fig. 1b) with a slowly varying effective RI slightly smaller than that of air. Moreover, the modal absorption and transmission losses show a clear presence of a photonic bandgap (see Fig. 1a) with material absorption being the dominant loss mechanism. At the same time, the predicted modal absorption loss in the center of the bandgap (~0.05 cm$^{-1}$) is considerably smaller than the PLA bulk absorption loss of ~0.95 cm$^{-1}$, thus indicating that efficient Bragg waveguides can still be realized while using high-loss materials that are typically employed in 3D printing.

The band diagram in Fig. 1b also features the dispersion relation of the reflector surface state that intersects the fundamental mode dispersion relation inside of the bandgap. The field distribution of this surface state, however, is mostly localized in the reflector region (see Fig. 1c II) and overlaps only slightly with that of the fundamental core-guided mode (see Fig. 1c I), thus being irrelevant for the Bragg waveguide operation.

*2.2 Hollow-core Bragg waveguide with a defect in the reflector*

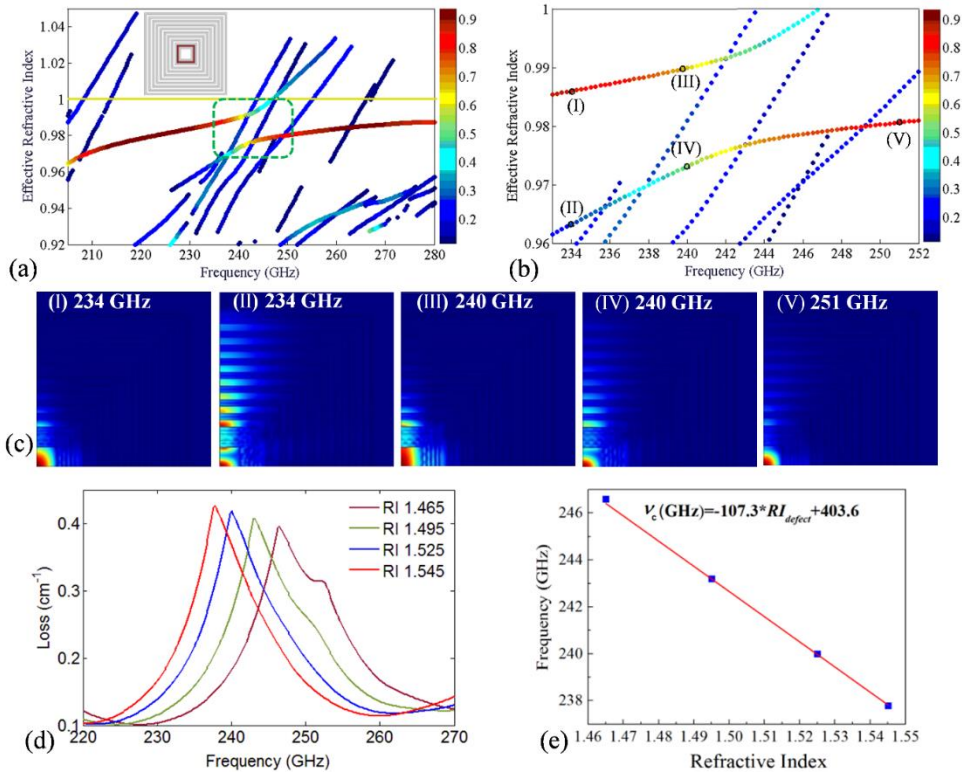

Fig. 2. (a) Band diagram of the hollow-core Bragg waveguide with a defect in the reflector second layer. The color scale defines the fraction of the modal power confined to the waveguide core. Only the modes with over 10% of power guided in the core are presented. Inset: schematic of the cross section of a Bragg waveguide with a defect. Red color indicates the position of the first air layer filled with liquid analyte. (b) A zoom into the spectral region of anticrossing between dispersion relations of the fundamental and the defect-guided modes. (c) Field distributions corresponding to the two hybridized modes in the anticrossing region (only a quarter of the computation cell is shown). Inserts I, V and II show the core mode and the defect mode at the edges of the anticrossing region, where modal hybridization is small. Inserts III and IV show the two hybridized modes at the center of the anticrossing region. (d) The total transmission loss of the fundamental (lowest loss) mode propagating through the Bragg waveguides with a defect, for various values of the defect RI. The frequency with the highest loss of the fundamental mode approximately corresponds to the frequency of phase matching between the core-guided and defect modes (mid-range of the anticrossing spectral region). (e) Relation between the RI of the defect layer and the frequency with the highest loss (anticrossing frequency). The red line is a linear fit of the numerical data. Theoretical sensitivity from the line slope is 107.3 GHz/RIU.

Next, we perform numerical studies of a Bragg waveguide with a defect in its reflector structure. Particularly, the RI of the air layer closest to the waveguide core is increased to 1.525, which corresponds to the RI of a typical analyte used in our experiments. Due to anticrossing of the core-guided mode and the defect state dispersion relations, the waveguide band diagram (Fig. 2a) is significantly different from the one for the waveguide with no defect (Fig. 1b). In the anticrossing region, the two modes are strongly hybridized with the supermode fields present both in the hollow core and in the defect layer (Fig. 2b and 2c III, IV), while at the edges of the anticrossing region, the supermodes have a predominant concentration of their fields either in the hollow core (Fig. 2b and 2c I, V) or in the defect layer (Fig. 2b and 2c II).

Moreover, transmission losses of the supermodes increase strongly inside of the anticrossing region as in this spectral region the low-loss hollow core-bound mode is strongly hybridized with a much lossy and tightly confined leaky mode that is bound to the defect layer. Experimentally, this leads to a sharp increase of the waveguide propagation losses inside of the anticrossing region. Theoretically, this phenomenon can be visualized by plotting spectral losses of the lowest-loss mode with effective RI smaller than one, which also happens to be the mode with the highest field presence in the waveguide core (see Fig. 2d). The resultant loss curve features a sharp loss peak whose spectral position corresponds to the center frequency of the anticrossing region (the frequency at which the difference between the two supermode propagation constants is smallest). Furthermore, similar loss curves can be plotted for various values of the defect RI (see Fig. 2d), from which it follows that spectral position of the loss peak is highly sensitive to the real value of the defect RI and it moves towards lower frequencies for higher values of the defect RI. When plotting the peak spectral position as a function of the defect layer RI, one finds a linear dependence between these two parameters in the whole operational range of the sensor (frequencies in the waveguide bandgap) (see Fig. 2e). The sensitivity of thus defined Bragg waveguide-based sensor is calculated as 107.3 GHz/RIU, while the spectral width of the loss peaks (defined as full width at the half maximum) is over 10 GHz, which is easy to resolve using both THz time domain spectrometer (TDS) and CW-THz setups.

## 3. Fabrication of the hollow-core THz Bragg waveguides using FDM and SLA techniques

Among various fabrication techniques, 3D printing technology has been proven to be an effective method for the fabrication of THz waveguides and guided wave components. A number of plastic THz optical components fabricated by SLA and FDM 3D printing technologies have been recently reported [21,24–28]. In the SLA, the photosensitive resin is cured layer by layer using spatially modulated UV radiation at the build plane. In the FDM, the plastic filament is first softened and then extruded through a nozzle as it is displaced in space to build the desired structure layer by layer. Within the FDM approach, the geometrical precision of the printed models as well as the roughness of its features are mainly decided by the nozzle size which is, typically, much larger (~200 – 400 μm) than the beam spot of the UV laser used in the SLA (10 – 50 μm) [38,39]. Therefore, when higher resolution (smaller roughness) is required, one normally resorts to SLA, while FDM is rather used for rapid prototyping of coarse models. At the same time, FDM single material print can be used as is, while SLA prints require removal of the unreacted photosensitive resin (normally using isopropanol), as well as final UV curing of the remaining prints. The postprocessing of SLA prints with solvents can cause significant problems due to surface tension effects when printing porous structures or long slender channels. As our waveguides contain multiple air channels in their reflector, we used both FDM and SLA techniques for their fabrication in order to compare these two methods.

Recently, our group has reported fabrication of small sections (2.5 cm in length) of circular hollow-core THz Bragg waveguides using SLA 3D printing [21]. In these waveguides, individual 512-μm thick resin layers were spaced with 512-μm thick air gaps that also contained two 145 μm by 512 μm rectangular spacers to provide mechanical stability to the whole reflector structure. The main problem in fabrication of such sections was post-printing elimination of the unreacted resin from the waveguide channels, as well as maintaining consistent layer thickness along the waveguide, which was proved to be difficult for sections longer than several cm.

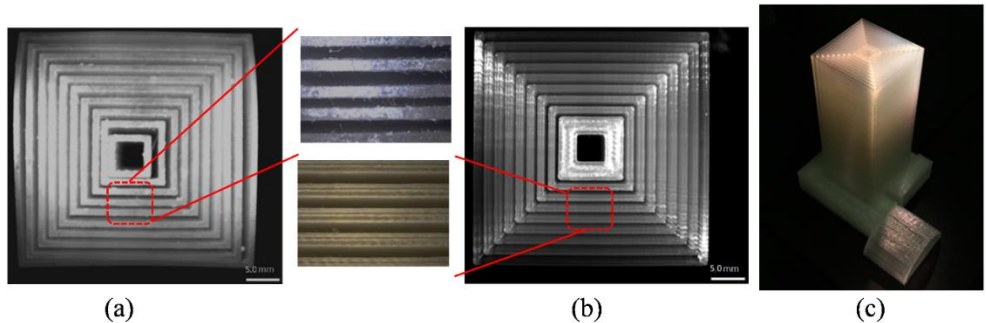

Fig. 3. (a) The cross section of the 3D-printed Bragg waveguide-based sensor fabricated using SLA and photosensitive resin. The structure is shown without the cover sheet that is used to seal the gap (defect) between the first and the second resin layers. (Dotted red region) The enlarged view of the reflector multilayers features layers of variable thickness and inconsistent periods. (b) The cross section of a 3D-printed Bragg waveguide-based sensor (FDM, PLA), also featuring a cover sheet that seals the air gap of a defect layer. (Dotted red region) The enlarged view of the reflector multilayers features a well-defined structure with consistent period and layer thickness. (c) A 3D-printed (FDM, PLA) hollow-core Bragg waveguide-based THz fluidic sensor that includes waveguide with a defect, and a fluidic delivery system.

In this work the main challenge was a single-shot fabrication of the whole waveguide structure that includes both the fluidic part and a relatively long (10 cm or more) waveguide. Such a long waveguide length is required to realize an effectively single mode guidance regime for which other higher-loss core-guided modes would radiate out before arriving at the waveguide end. Inspired by our prior work, we first attempted a 10-cm long Bragg waveguide fabrication using the SLA technique (ASIGA Pro 2 with PlasClear photosensitive resin) with the result shown in Fig. 3a. For comparison, we also used FDM with PLA filament (MakerBot Replicator) to fabricate the same structure (see Fig. 3b and 3c).

When comparing the two prints we observe that the resultant structure of the waveguide fabricated using SLA is inferior to the one fabricated using FDM mainly due to the resin being much softer and pliable compared to PLA. As a result, the relative position of the printed resin layers separated by air gaps is easily affected by even small internal stresses during manufacturing and postprocessing, as well as external stresses when handling. In addition, as a postprocessing, we used the isopropanol washing followed by the airflow drying to remove the uncured resin that fills voids in the fabricated model. The uncontrollable effects of the surface tension during washing as well as apparent swelling of the resin when exposed to the washing agent have a significant effect on the internal porous structure of a long Bragg waveguide, thus resulting in pronounced dislocations and deformations of the individual resin layers. As observed under the microscope, the thicknesses of the resin layers and the air gaps in a cleaned structure vary between 0.7 and 1.0 mm, whereas both layer thickness in the printed structure before resin removal is around the designed 0.8 mm. Therefore, the standard cleaning process with isopropanol used in SLA printing results in a high degree of structural nonuniformity in long and highly porous Bragg waveguides studied in this work. As a result, due to significant variations in the geometry of the printed Bragg reflector from the theoretical one, SLA printed waveguides after cleaning show relatively small and shallow bandgaps, while anticrossing resonant phenomenon is not observed at all as it requires precise geometry for the defect layer and the Bragg reflector. Finally, due to the challenges associated with extraction of the unreacted resin from the waveguide porous structure, the sensor could not be printed as a single piece and it requires post printing assembly of at least four components (the waveguide, fluidic conduits and a cover sheet).

Next, we have attempted a single-step fabrication of the Bragg waveguide-based fluidic sensors using FDM technique with a PLA filament. Compared to the Asiga Pro 2 SLA printer that offers z-axis resolution of 10 μm and x-y axis resolution of 50 μm, the FDM Makerbot Replicator printer offers 100 μm resolution along the z axis and 400 μm resolution along the x-y axis. At the same time, the PLA material used in FDM is more rigid than the resin used in SLA, and the whole sensor structure can be printed in a single run without the need of any post processing. As a result, while the structures created using FDM feature rougher surfaces and potentially higher scattering losses, at the same time, relative rigidity of the PLA material and the lack of post processing result in highly repeatable Bragg waveguide-based sensors with well-defined geometries (see Fig. 3b). Particularly, we find that the FDM Bragg reflectors consistently feature PLA layer and air gap thicknesses in the range of 0.75 – 0.85 mm compared to the designed 0.8 mm. Additionally, unlike the corresponding SLA fabricated structures, the FDM fabricated hollow-core Bragg waveguides feature deeper bandgaps and anticrossing resonant phenomenon when liquid analyte is placed into the first air gap.

Theoretically, the linewidth of the absorption dip caused by the anticrossing phenomenon is decided by the degree of the field overlap between the core-guided and defect modes. In principle, to achieve higher sensing resolution, it is desirable to reduce the spectral width of the absorption dip, which can be achieved by placing the defect in the Bragg reflector (liquid analyte-filled air gap) further away from the hollow core [21]. Thus, according to our simulations, when the defect layer is moved from the first to the second air gap, the spectral width of the loss peak (defined as full width at the half maximum) induced by the anticrossing resonant phenomenon decreases from ~10 GHz to ~5 GHz, while further reduction in the peak width is possible by moving the defect layer farther away from the core. Experimentally, while the absorption dip is consistently and readily observable with a defect in the first layer, when moving the defect into the air gaps farther away from the core, it becomes harder to detect due to considerably less pronounced intensity variation. The decrease in the dip strength when placed into the farther air gaps can be explained by the variation in the reflector layer thicknesses and the scattering loss caused by the layer roughness, which are both known to reduce spatial coherence of the standing waves in the Bragg reflector that is needed for manifestation of any resonant phenomenon (see detailed discussion of this phenomenon in [40]). These decoherence effects are more pronounced for sharper resonances that are expected when placing the defect layer farther away from the waveguide core.

## 4. Characterizing of the hollow-core THz Bragg waveguide-based sensors using CW-THz spectroscopy

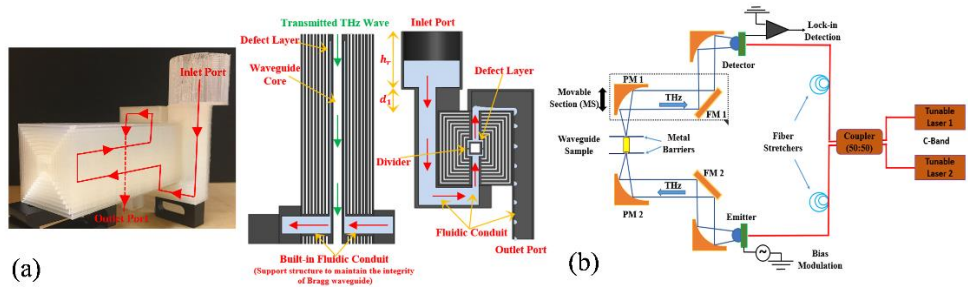

Fig. 4. (a) The fabricated hollow-core THz Bragg waveguide-based fluidic sensor and its schematics. In grey color — PLA material, in blue color — the liquid analyte, in white — air. Red arrows show the flow direction of the liquid analyte.

Green arrows show the direction of THz propagation. (b) Schematic of the CW-THz spectroscopy system used to characterize the sensors.

In what follows, we study the hollow-core THz Bragg waveguide-based fluidic sensor with a defect in the first air gap of the Bragg reflector (analyte-filled air gap). In order to place the liquid analyte into the defect layer, we have sealed the first air gap on both sides of a waveguide, while a fluidic conduit was placed into the gap to inject and maintain the flow of the liquid analyte through the channel (see Fig. 4a). In our experiments, different analytes were injected into the reflector gap, while optical measurements were performed in a static regime (no flow) with the defect layer being fully and slowly filled with the injected analyte starting from an empty system. The path of the liquid flow in the waveguide-based sensor is shown in Fig. 4a. We have also verified in a separate experiment that the fluid occupied completely the gap space by using colorants and dissecting the model after filling and then draining the sensor structure.

Both the transmitted amplitude and phase spectra of the fabricated THz Bragg waveguide-based sensor (Fig. 3c) were studied by using the CW-THz spectroscopy system (TOPTICA Photonics) [41,42]. In such a system (see Fig. 4b), we mix in a fiber radiation from two distributed feedback (DFB) lasers with slightly different emission wavelengths and combined power of ~30 mW. A 50:50 fiber coupler is used to split the signal into detection and emission arms. The THz radiation is generated in the photomixer (under bias voltage) and corresponds to the beat frequency between the two DFB lasers. A similar photomixer (without bias) and a lock-in amplifier are used to detect the THz radiation. Reflective optics is used to focus a CW THz beam onto the input facet of a hollow-core Bragg waveguide, while the transmitted THz beam is re-collimated and detected using lock-in amplification. For CW-THz spectroscopy, the THz frequency is scanned by continuously varying the emission wavelengths of the two DFB lasers via adjusting the temperature/current of the Bragg grating in each laser cavity. Additionally, at every frequency, a phase measurement is performed by using an optical delay line in the form of polarization-maintaining fiber wound piezoelectric elements. Finally, the optical paths are adjusted before insertion of a Bragg waveguide to result in a flat phase response for the reference spectrum (empty system without a waveguide) [43].

Transmitted THz spectra of the hollow-core THz Bragg waveguide sensors are measured in the following processes. First, the reference spectrum is recorded by placing a metallic aperture with an opening identical to that of the waveguide core at the joint focal point of the two parabolic mirrors PM1 and PM2 (see Fig. 4b). Note that PM1 and a flat mirror FM1 are both mounted onto a rail, and represent a movable section (MS) of the spectrometer, thus allowing measuring waveguides of various sizes, as well as a reference signal without a waveguide. Next, the Bragg waveguide is inserted with its input and output facets at the focal points of PM1 and PM2 via adjusting the MS position. Additionally, two metallic apertures with opening identical to that of a hollow core are placed adjacent to each facet in order to block excitation of the cladding modes. Finally, the waveguide transmission amplitude and phase are extracted by comparing the measurements with and without (reference) a waveguide.

*4.1 Choosing the optimal length of the hollow-core Bragg waveguide*

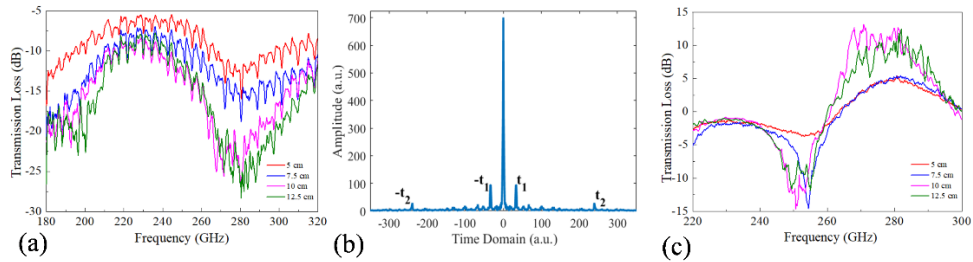

Fig. 5. (a) Normalized transmission spectra of the FDM fabricated hollow-core Bragg waveguides (no defect) featuring different lengths. Normalization is with respect to the transmission spectra of the empty system. (b) Fourier transform of the measured transmission spectrum through the empty spectrometer (reference). (c) Resonant absorption dip (anticrossing phenomenon between the defect state and a core-guided mode) when filling the first air gap of the Bragg waveguides with liquid analyte. Different transmission spectra are for waveguides (with defect) of different lengths. Normalization is with respect to the transmission spectra of a waveguide of the same length without a defect (no analyte).

Transmission spectra of the hollow-core Bragg waveguides depend strongly on the waveguide length. This is related to the well-known fact that large hollow-cores of most PBG waveguides support a large number of leaky modes that feature greatly different propagation losses. Thus, by choosing a long enough waveguide so that only the lowest-loss (fundamental) mode remains at the waveguide output end, one observes the well-defined transmission bandgaps. In contrast, for shorter waveguides, contribution of the higher-loss modes remains important, which results in transmission bandgaps that tend to broaden and become shallow. Clearly though, when the waveguide is too long, transmission loss of the fundamental leaky mode can be significant, thus resulting in the low signal to noise ratio. Therefore, ideally, the hollow-core Bragg waveguide has to be long enough to result in the effective single mode guidance, while not too long in order to avoid excessive transmission loss. Therefore, we first studied the optimal size of the hollow-core THz Bragg waveguide to be used in our sensors.

In Fig. 5a, we plot the normalized transmission loss spectra (by field) for the PBG Bragg waveguides with the lengths of 5 cm, 7.5 cm, 10 cm and 12.5 cm as:

$$Loss(\upsilon) = 10 \cdot \log 10(T_{waveguide}(\upsilon)/T_{ref}(\upsilon)) \quad (1)$$

From this data, we see that the spectral size and the depth of bandgap stabilize for waveguides of 10 cm and longer.

Furthermore, in the transmission data, we also see intensity ripples superimposed onto the waveguide transmission spectra. In fact, there are two sets of ripples, one is caused by the multiple reflections inside of the two silicon lenses (size $d$~10 mm, $n_{Si}$~3.42, [44]) that are integrated into the THz photomixers in the emission and detection arms of our setup. The corresponding frequency of these ripples is

$$\Delta \upsilon_1 = \frac{c}{2dn_{si}} \sim 4 \text{ GHz} \quad (2)$$

Another set of ripples with higher frequencies is due to standing waves excited in the spectrometer cavity of length $L \approx 80$ cm (distance between the detector and emitter) with the characteristic frequency of

$$\Delta \upsilon_2 = \frac{c}{2Ln_{cavity}} \sim 0.2 \text{ GHz} \quad (3)$$

In Fig. 5b, we present Fourier transform into time domain $T(t)$ of the measured transmission spectra $T_{ref}(\nu)$ of an empty system (reference):

$$T_{ref}(t) = \int T_{ref}(\upsilon) * \exp(-i\upsilon t) d\upsilon \quad (4)$$

$$T_{ref}(\upsilon) = \int T_{ref}(t) * \exp(i\upsilon t) dt \quad (5)$$

The FFT shows clearly the presence of five peaks, one at $t=0$, and four others at $\pm t_1$ and $\pm t_2$ respectively. This means that the transmission spectrum contains among other frequencies the two sinusoidal contributions that define spectral ripples as:

$$T_{ref}(\upsilon) = T_0 + T_1 \cos(\upsilon t_1 + \varphi_1) + T_2 \cos(\upsilon t_2 + \varphi_2) + \cdots \quad (6)$$

From Eq. (6) we thus find two characteristic ripple frequencies, one associated with standing waves inside silicon lenses as:

$$\Delta \upsilon_1 = \frac{2\pi}{t_1} \sim 4 \text{ GHz} \quad (7)$$

while another one corresponding to the standing waves in the spectrometer cavity as:

$$\Delta \upsilon_2 = \frac{2\pi}{t_2} \sim 0.2 \text{ GHz} \quad (8)$$

Although the intensity of these spectral oscillations can be somewhat mitigated by normalizing with respect to the reference, even the normalized data features spectral ripples that ultimately limit the sensor resolution.

Next, we study the shape and the depth of an absorption dip as a function of the Bragg waveguide length when introducing a defect into the structure of the Bragg reflector. In Fig. 5c, we present transmission spectra through waveguides of different lengths with the first air gap filled with mineral oil of $n_{oil}$~1.455. The spectra are normalized with respect to the transmission spectra of the corresponding waveguides but without analyte (empty sensor). Similarly to the case of a waveguide without defect (Fig. 5a), we observe that the shape of the resonance, its depth and spectral position, stabilize for the waveguides longer than 10 cm.

Finally, we note that the volume of the analyte filling the first gap is estimated at ~2 mL.

## 5. The performance of the THz PBG Bragg waveguide-based fluidic sensor

In this section, we characterize sensitivities of our THz fluidic sensors using mixtures of oils with RI in the range between 1.465 and 1.545 as liquid analytes. Firstly, the RIs of the mineral oil (Johnson & Johnson Inc) and cinnamon essential oil (Simply Essentials, Walmart Inc) were measured as 1.455 and 1.555 using a THz time-domain spectrometer system in a cutback configuration (see Supplementary Material). Then, nine sets of liquid analytes with RI of 1.465, 1.475, 1.485, 1.495, 1.505, 1.515, 1.525, 1.535, and 1.545 were prepared by mixing thoroughly these two kinds of oils with volume ratios of 9:1, 8:2, 7:3, 6:4, 5:5, 4:6, 3:7, 2:8, and 1:9 respectively. Finally, two sets of measurements were performed to characterize sensor sensitivity, one using amplitude detection modality and another using phase detection modality.

*5.1 Amplitude detection modality*

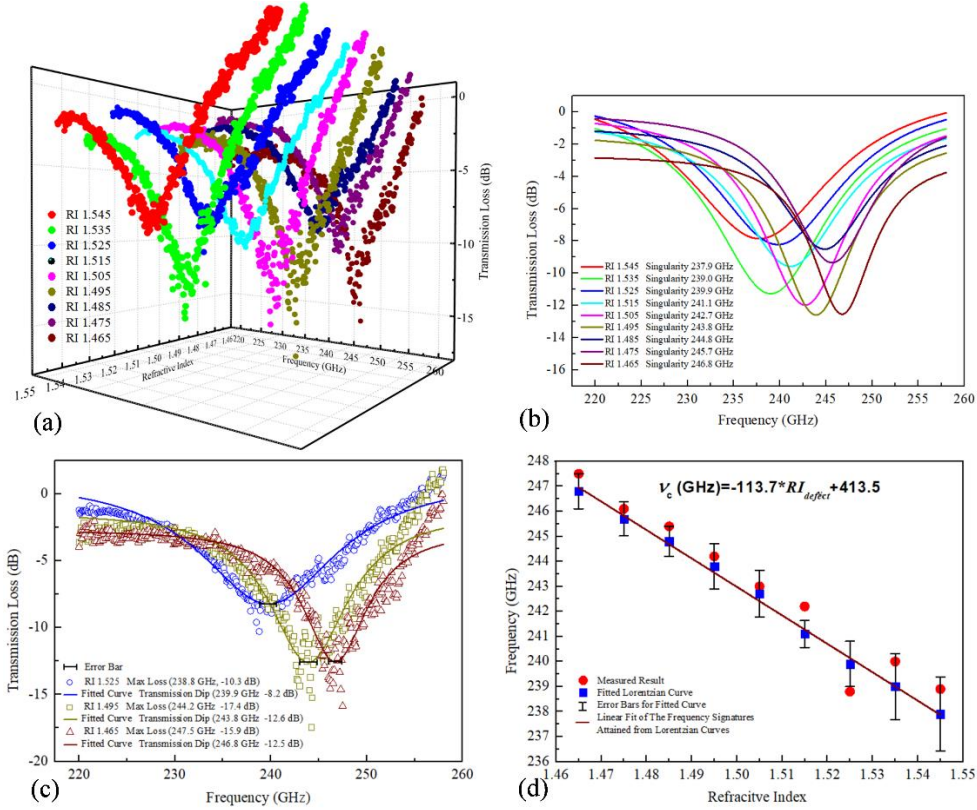

Fig. 6. (a) Normalized transmission in the vicinity of absorption peak for different values of the analyte RI. (b) Lorentzian fit of the experimental curves. (c) Estimation of error in the position of a spectral dip using mean square deviation of the Lorentzian fit and the raw data; shown are examples for the analyte refractive indices of 1.525, 1.495, and 1.465. (d) Dependence of the spectral position of the absorption dip on the analyte RI. Blue dots and error bars are computed using Lorentzian fitting curve and mean square error estimation. Red dots correspond to the maxima of the transmission losses.

In the first experiment, sensor sensitivity was deduced from the spectral position of the absorption dip as detected from the normalized transmission spectra (by field) of the Bragg waveguide with analyte-filled defect. To obtain a normalized spectrum, we have divided transmission spectrum of a waveguide with an analyte-filled gap (defect) by that of an empty waveguide (no analyte). In our measurements, we used 100 MHz frequency resolution. In Fig. 6a, the raw normalized transmission data is presented in a 3D plot as:

$$T(\upsilon) = 10 \cdot \log10(T_{analyte}(\upsilon)/T_{empty}(\upsilon)) \quad (9)$$

In Fig. 6b, we present analytical fitting of the raw data using Lorentzian lineshapes:

$$T_L(\upsilon, \upsilon_c, \Delta\upsilon, A, T_0) = T_0 + A\frac{\Delta\upsilon}{4(\upsilon-\upsilon_c)^2 + \Delta\upsilon^2} \quad (10)$$

This is accomplished by minimizing the weighting function defined as a mean square error of the fit:

$$MSE(\upsilon_c, \Delta\upsilon, A, T_0) = \sum_{\upsilon_{min}}^{\upsilon_{max}} (T_L(\upsilon) - T_{exp}(\upsilon))^2 \quad (11)$$

Using data fitting with Lorentzian lineshapes allows efficient filtering of the spectral ripples from the raw normalized data, as well as to deduce spectral position $\upsilon_c$ and spectral width $\Delta\upsilon$ of the absorption dip. While spectral position of the dip is sensitive to the analyte RI, its spectral widths remains almost constant $\Delta\upsilon \sim 12$ GHz.

Furthermore, comparing the raw data with its Lorentzian fit (Fig. 6c) also allows us to define the error in the definition of the absorption dip position. Numerically, the errors in the spectral dip position ($\pm\delta\upsilon_c$) can be found using the following definitions:

$$MSE'(v_c, \Delta v, A, T_0) = \sum_{v_C - dip\ width/2}^{v_C + dip\ width/2} (T_L(v) - T_{\exp}(v))^2 \tag{12}$$

$$MSE'(v_c \pm \delta v_C, \Delta v, A, T_0) = (e-1) \cdot MSE'(v_c, \Delta v, A, T_0) \tag{13}$$

The values of the absorption peak position (blue dots) as a function of the analyte RI, and the corresponding error bars are presented in Fig. 6d, from which we observe that absorption dip moves to lower frequencies with increasing analyte RI. The average error in the peak position is $\delta v_c$ ~0.9 GHz, which is attributed to the noise due to standing wave resonances in the CW-THz spectroscopy system as described earlier in the paper. Additionally, in Fig. 6d, for each value of the analyte RI, we plot the frequency of highest transmission loss (red dots) which can be also used as an alternative definition of the spectral position of the absorption dip. We note that the red dots generally fall within the error bars of the blue dots, thus indicating that the definition of the spectral position of an absorption dip based on fitting to Lorentzian lineshape is comparable in accuracy to the definition based on the highest transmission loss.

Finally, from Fig. 6d, we conclude that the sensor response is linear with the value of the analyte RI with the corresponding sensitivity of $S$=113.7 GHz/RIU. Considering the ~50 GHz bandgap bandwidth and the ~12 GHz linewidth of the absorption dip, our sensor is able to detect liquid analytes with RI in the range between 1.4 and 1.8, which covers most of the common organic solvents and liquids for chemical and biological applications [45,46]. Sensor resolution using amplitude detection modality can be estimated from the average value of the error in the absorption peak position as:

$$\delta n = \delta v_c / S \sim 8.0 \cdot 10^{-3} \text{ RIU} \tag{14}$$

*5.2 Phase detection modality*

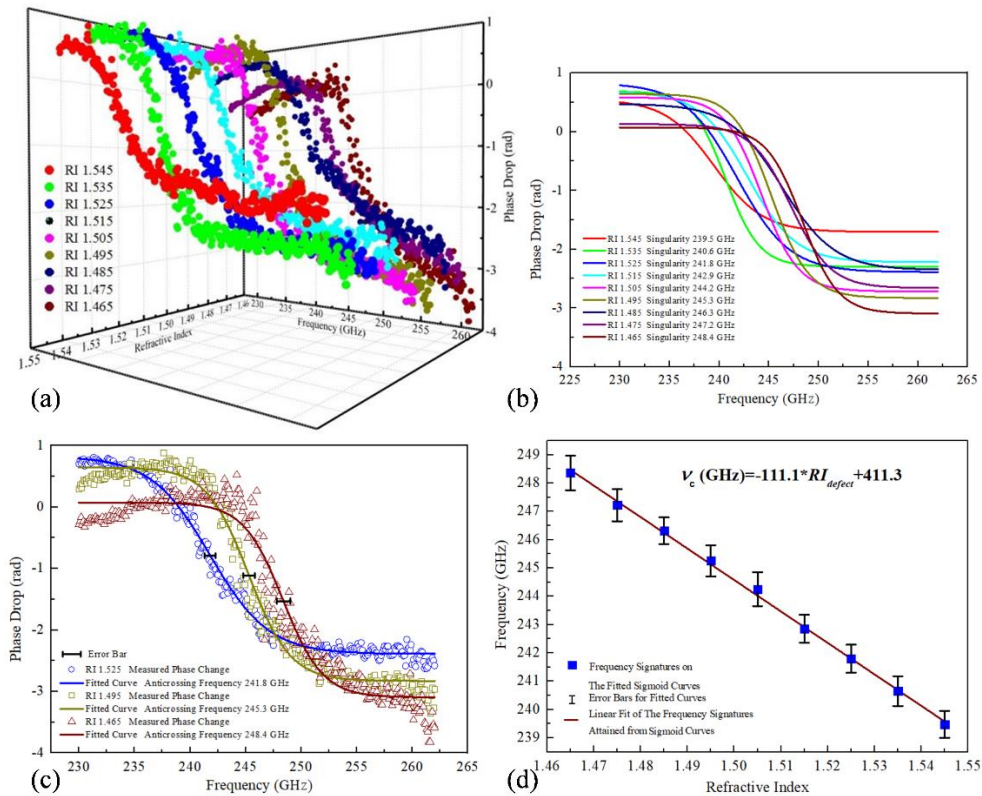

Fig. 7. (a) Phase dependence extracted from the complex normalized transmission in the vicinity of the absorption peak (anticrossing frequency) for analytes featuring different values of RIs. (b) Sigmoid fit of the experimental curves. (c) Estimation of error in the position of the anticrossing frequency using mean square deviation of the Sigmoid fit and the raw data; shown are examples for the analyte refractive indices of 1.525, 1.495, and 1.465. (d) Dependence of the anticrossing frequency on the analyte RI. Blue dots and error bars are computed using Sigmoid fitting curve and a mean square error estimation.

In the second experiment, sensor sensitivity was deduced from the spectral position of the anticrossing frequency between the core mode of a Bragg waveguide and a mode of an analyte-filled defect as deduced from the rapid change in the phase of a transmitted light. The phase was retrieved from the complex normalized transmission spectrum (by field) obtained by dividing the transmission spectrum of a waveguide with an analyte-filled gap (defect) by that of an empty waveguide (no analyte). In our measurements, we used 100 MHz frequency resolution. In Fig. 7a, the raw phase data is presented in a 3D plot. In Fig. 7b, we present analytical fitting of the raw data using sigmoid lineshape:

$$\varphi(\upsilon,\upsilon_c,\Delta\upsilon,\varphi_0,\Delta\varphi) = \varphi_0 + \frac{\Delta\varphi}{1+\exp(\frac{\upsilon-\upsilon_c}{\Delta\upsilon})^2} \quad (15)$$

which allows us to deduce the anticrossing frequency $\nu_c$, as well as the spectral width of the anticrossing region $\Delta\nu$. This is accomplished by minimizing the weighting function defined as a mean square error of the fit similarly to Eq. (11). Furthermore, similarly to Eq. (12) and (13), comparing the raw data with its fits using sigmoid lineshapes (Fig. 7c) allows us to define the error $\pm\delta\nu_c$ in the position of the anticrossing frequency. Spectral bandwidth of the anticrossing region is found to be weakly dependent on the analyte RI and equal to $\Delta\nu \sim 2$ GHz, which is significantly smaller than the 12GHz-spectral width of a resonant dip as defined using the amplitude-based methodology. In Fig. 7d, we present the value of the anticrossing frequency (blue dots) as a function of the analyte RI, and the corresponding error bars. The average error in the anticrossing frequency is found to be $\delta\nu_c \sim 0.5$ GHz, which is almost twice as small as in the case of amplitude-based detection described in the previous section. We attribute this to the fact that negative effects of the standing wave oscillations in a CW-THz system are more efficiently mitigated when working with a normalized signal phase rather than a normalized signal amplitude. This is also confirmed by comparing the quality of the raw data and its fits using the amplitude-based detection (Fig. 6c) with those of the phase-based detection (Fig. 7c), with phase-based approach showing considerably smaller data spread and fit errors than the amplitude-based one.

Finally, from Fig. 7d we conclude that the sensor response is linear with the value of the analyte RI with the corresponding sensitivity of $S$=111.1 GHz/RIU, which is almost identical to that of the amplitude-based detection method. At the same time, sensor resolution using phase detection modality is superior to that of the amplitude detection modality and can be estimated as:

$$\delta n = \delta\upsilon_c / S \sim 4.5 \cdot 10^{-3} \text{ RIU} \quad (16)$$

### 5.3 Real time RI monitoring using fluidic sensor

In this section we discuss applications of the proposed resonant fluidic sensor in real time monitoring of the fluid RI in motion. All the measurements presented so far were conducted in a static regime in which different analytes (oils with calibrated RIs) were slowly loaded into an empty sensor. Moreover, we have ensured that the analytes occupied all the volume in a defect layer by referencing it to the volume of the used oil. Additionally, we have dissected several waveguides after following this filling procedure using oil with a colorant and observed that the inner surfaces of all the conduits and a sensing layer were colored. When using fluidic sensors for real time monitoring of flowing liquids, one has to ensure that the analyte liquid injected from the inlet opening flows through the whole sensing layer channel before arriving at the outlet. This can be accomplished by placing into the channel two square dividers of sizes equal to that of the air gap and of lengths slightly smaller than the waveguide length in order to force the fluid to go through the whole length of a waveguide and back, rather than choosing the shortest distance between the inlet and outlet openings of a sensing layer (see Fig. 4a). In this design, the total effective length of a fluidic channel in a sensing layer (along the x axis) is $L_c$=17.4 cm, which is almost twice as long as the waveguide length, while the effective channel width $W_c$=1.3 cm is half of the perimeter of a sensing layer. Separation between the parallel plates $D_c$=0.8 mm equals to that of the air gap size. The sensor channel volume is then $V_c=L_c \cdot W_c \cdot D_c \sim 1.81$ cm$^3$. The sensor is fed from a semi-cylinder-shaped fluid reservoir of volume $V_r$=10.4 cm$^3$ (height $h_r$=2.6 cm, base area $A_r$=4 cm$^2$), which is connected to the inlet opening of a sensing channel via a square cross-section conduit (0.84 cm × 0.84 cm) with a total dead volume of $V_d$=5 cm$^3$. A section of this channel of height $d_1$=1.1 cm is located below the reservoir but above the level of the output fluidic port.

In the following experiments, we used the hydrostatic pressure of the fluid column of the total height $h+d_1$ ($h$ is the height of liquid in the reservoir $0<h<h_r$) to drive the analyte flow through the sensing layer. The density of the liquid analyte (Blandol and Carnation white oil mixture) is measured as $\rho$=857 kg/m$^3$ and viscosity is $\mu$=0.012 Pa·s.

The flow is purely gravity driven with hydrostatic pressure $P(h)=\rho g(h+d_1)$ and assumed as laminar and viscosity-driven ($Re<<1$), we use the well-known expression for fluid velocity profile of creeping flow in parallel plate geometry (along the y direction) [47]:

$$v(y,h) = \frac{\rho g (h+d_1)}{2\mu L_c} y(D_c - y) \tag{17}$$

Further using the electric-hydraulic analogy [48], we define a pressure - flow rate relationship in the form of $P(h)=R_{hyd} \cdot Q(h)$, where

$$R_{hyd} = \frac{12\mu}{D_c^3} \cdot \frac{L_c}{W_c} \tag{18}$$

is the channel hydraulic resistance, which for our system is estimated as $R_{hyd}^{th}$~3.8 kPa·s/mL. Assuming that the liquid is incompressible, and using Eq. (18) along with a mass balance between the reservoir and main channel, we can find the dependence of the liquid height in the reservoir as a function of time:

$$-A_r \frac{dh}{dt} = Q(h) = \frac{\rho g (h+d_1)}{R_{hyd}} \tag{19}$$

Furthermore, assuming that the reservoir is completely filled at the beginning of the measurement, Eq. (19) has the following exponential solution:

$$h(t) = (h_r + d_1) \cdot \exp(-\frac{\rho g}{A_r R_{hyd}} t) - d_1 \tag{20}$$

$$Q(h) = \frac{\rho g (h_r + d_1)}{R_{hyd}} \cdot \exp(-\frac{\rho g}{A_r R_{hyd}} t) \tag{21}$$

that are valid for $0<h(t)<h_r$. From Eq. (20) we can relate the reservoir voiding time $t_r$ to the channel hydraulic resistance $R_{hyd}$ as:

$$h(t_r) = 0 \rightarrow t_r = \frac{A_r R_{hyd}}{\rho g} \cdot \ln(1+\frac{h_r}{d_1}) \tag{22}$$

Using the values of geometrical and material parameters presented above, we now find the theoretical estimate of the reservoir voiding time $t_r$=211 s.

Additionally, when operating a sensor with a fully filled reservoir, the filling time of a sensing layer can be estimated as:

$$t_s = \frac{R_{hyd} V_c}{\rho g (h_0 + d_1)} = \frac{12\mu}{\rho g (h_0 + d_1)} \cdot (\frac{L_c}{D_c})^2 \sim 21 \text{ s} \tag{23}$$

which is also comparable to the scanning time ~10—30 s (depending on the resolution and averaging constant) of our CW spectroscopy system over the spectral range covered by the Bragg waveguide bandgap. Note that the filling time of a sensing layer as defined by Eq. (23) does not depend on the channel width but rather on the ratio of the channel length to the channel size. This time can also be considered as a characteristic response time of the sensor under continuous non-forced (no external pressure) liquid flow. This means that if the RI of an analyte at the input port of a sensing channel is suddenly changed to another value, the sensor spectral response will stabilize to a new reading after ~$t_s$. Experimentally, sensor response time can be measured by simply introducing a new analyte into the reservoir and then marking the time it takes the sensor to stabilize to a new reading after changes start appearing in the sensor spectral response.

Finally, the maximum Reynolds number $Re=\rho U L_{ch}/\mu$ for the liquid flow within the sensing channel can be computed. We use the highest possible flow rate value (full reservoir) to compute the characteristic velocity $U \sim Q_{max} L_c / V_c$ in conjunction with the well-known characteristic length $L_{ch} \sim D_c^2/L_c$ for creeping flows in elongated channel [48]. For the current design, the maximum Reynolds number within the sensing channel gives:

$$Re_{max} = \frac{\rho \frac{Q_{max}}{V_c} \cdot L_c \cdot \frac{D_c^2}{L_c}}{\mu} \sim 0.002 \tag{24}$$

which validates our initial creeping flow assumption.

Next, we experimentally measure the voiding time of the analyte reservoir, as well as sensor response time in order to verify if the channel volume used by the liquid flow corresponds to that of the physical volume of a sensor. This is important as sensor response time and its sensitivity will be directly affected by the speed and extent to which a new analyte could substitute the old analyte already present in the waveguide structure. In particular, we compare performance of the two sensors, one using a divider as shown in Fig. 4(a), while another without a divider. While both sensors have virtually identical optical response in the static regime (no liquid flow during measurement), they behave rather distinctly in the dynamic regime.

For a sensor without a divider the reservoir voiding time is found to be $t_r^{no\ div} \sim 30$ s, which is seven times smaller than the theoretical one $t_r=211$ s. Furthermore, using experimentally measured value of the reservoir voiding time together with Eq. (22) we can find the channel hydraulic resistance $R_{hyd}^{no\ div} \sim 0.5$ kPa·s/mL, which is much smaller than the value derived from the theoretical analysis $R_{hyd}^{th} \sim 3.8$ kPa·s/mL. From the definition of the $R_{hyd}$ we find that for a channel without a divider the effective channel aspect ratio $(L_c/W_c)^{no\ div} \sim 1.9$, which is much smaller than the geometrical value of $L_c/W_c \sim 13.4$. This indicates that in a sensor without a divider the liquid flows along the shortest path from the inlet to the outlet mostly perpendicularly to the waveguide rather than along its length. As a result, when attempting measurement of the sensor dynamic response, instead of the theoretically predicted $t_s=21$ s, it takes the sensor without a divider $t_s^{no\ div} \sim 3$–5 min to stabilize to a new value indicating that the flow of a new analyte does not efficiently displace the old analyte from the sensing channel.

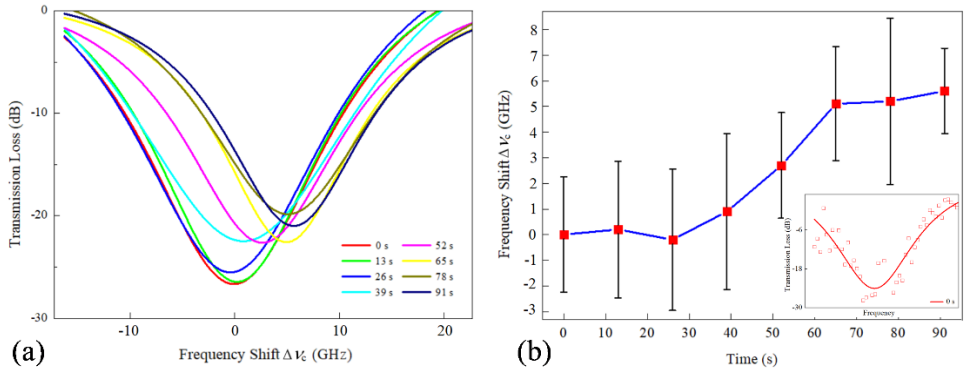

Fig. 8. (a) Lorentzian fits of the time dependent normalized transmission spectra of a sensor during displacement of an old analyte with RI of 1.505 with a new one with RI of 1.455 in the sensing layer. (b) Corresponding time dependence of the spectral position of the absorption dip. Error bars are computed by using mean square deviation of the Lorentzian fits of the raw data. Insert: example of the raw data and the corresponding Lorentzian fit at 0 s.

In contrast, for a sensor with a divider, the voiding time of the reservoir is measured as $t_r^{div} \sim 195$ s, while channel hydraulic resistance is found as $R_{hyd}^{div} \sim 3.4$ kPa·s/mL, which both agree closely with the theoretically estimated ones. Finally, when performing dynamic spectral measurement by introducing the analyte of a distinct RI into the sensor reservoir, it takes $t_s^{div} \sim 30$ s for the sensor spectral response to stabilize (see Fig. 8), which agrees well with the theoretically predicted value of $t_s=21$ s. A somewhat larger value of the sensor response time can be due to the inherent latency of the CW spectroscopy system that uses lock-in acquisition and data integration can add additional 10—30 s to the response time depending on the spectral resolution of a scan. Additionally, further optimisation of the fluidic design is necessary to make sensor operation more consistent and reliable. Namely, due to very low rates of the gravitationally driven fluid flow, the output port of a relatively large size (0.84 cm × 0.84 cm) is largely empty (see Fig. 4(a) right) with droplets forming on the port inner walls. Therefore, the hydrostatic pressure taken as $\rho g(h+d_1)$ can be somewhat different from this value as $d_1$ parameter can vary by several mm due to dynamic droplet formation and the fluid outlet. As $d_1$ parameter is used to relate the experimentally measured reservoir voiding time to the channel resistance, we expect that the values of the channel resistance and estimates for the channel filling times can also be affected.

## 6. Conclusion

In this work, a 3D printed THz Bragg waveguide-based resonant fluidic sensor has been investigated theoretically and experimentally. The sensor features a sealed fluidic channel placed into the reflector of a 10cm-long hollow-core Bragg waveguide, and it is fabricated in a single step using a low-resolution (400 μm in X-Y, 200 μm in Z) FDM printer and a PLA filament. Then, a CW-THz spectroscopy in the 0.1-0.3 THz spectral range is used to characterize sensitivity and resolution of sensor via tracking of the spectral position of the resonant absorption dip as a function of the RI of analyte in the fluidic channel. Both amplitude detection and phase detection methods are used for monitoring of the absorption dip position, with phase detection method resulting in higher quality data and lower noise. The spectral position of the transmission dip is found to depend linearly on the analyte RI, while moving towards lower frequencies for higher values of the analyte RI. The sensitivity of our THz Bragg waveguide-based fluidic sensor is measured to be ~110 GHz/RIU, which also matches well the theoretical simulations. Finally, sensor resolution using phase detection modality is estimated to be $\delta n$ ~4.5·10$^{-3}$ RIU and is mostly due to the negative effect of the standing waves in various optic elements of the CW-THz spectrometer cavity. We believe that due to the ease of fabrication and good optical characteristics, this proposed fluidic sensor can find its applications in real-time monitoring of the analyte RI.

## Funding

Canada Research Chairs program for their financial support of the Prof. Skorobogatiy Chair I in "Ubiquitous THz photonics", as well as China Scholarship Council funding of the doctoral fellowship of Y. Cao.

# 3D-printed Terahertz Photonic Bandgap Waveguide-based Fluidic Sensor: Supplementary Material

YANG CAO,[1] KATHIRVEL NALLAPPAN,[1,2] HICHEM GUERBOUKHA,[1] THOMAS GERVAIS,[1] AND MAKSIM SKOROBOGATIY[1,*]

[1]*Engineering Physics, Polytechnique de Montréal, C. P. 6079, succ. Centre-ville, Montréal, Quebec H3C 3A7, Canada*
[2]*Electrical Engineering, Polytechnique de Montréal, C. P. 6079, succ. Centre-ville, Montréal, Quebec H3C 3A7, Canada*
*\*maksim.skorobogatiy@polymtl.ca*

**Abstract:** This document provides supplementary information to "3D-printed terahertz photonic bandgap waveguide-based fluidic sensor". In this document, we present the measurement of the refractive indices of materials employed in the main paper.

## 7. The RI of PLA used in FDM 3D printing

The RI and absorption loss of the PLA material used in the PBG Bragg waveguide fabrication are measured by employing a cutback method and a CW-THz spectroscopy. As shown in Fig. S1a, the printed PLA plates with thickness ~4 mm are first placed in a stack and then inserted in the path of a collimated THz beam. Then, several consecutive measurements of the transmitted complex field as a function of frequency are performed by removing the plates from the stack one after another. The RI and loss can then be extracted from the amplitudes and phases of the normalized transmission spectra by using transmission through an empty system as a reference (see Fig. S1b and S1c). Particularly, for the stack of length $L$ and complex refractive index $n(\omega)$, the following equations are used to interpret the measured data [1,2]:

$$T(\omega, L) = \frac{E_{stack}}{E_{ref}} = C_{in}(\omega) \cdot C_{out}(\omega) \cdot \exp\left(-\frac{\alpha(\omega)}{2} \cdot L\right) \cdot \exp(i\varphi(\omega, L)) \quad \text{(S1)}$$

$$\varphi(\omega, L) = -\frac{\omega}{c}(\text{Re}(n(\omega)) - 1)L \quad \text{(S2)}$$

$$\alpha(\omega) = \frac{2\omega}{c}\text{Im}(n(\omega)) \quad \text{(S3)}$$

where $E_{stack}$ and $E_{ref}$ are the complex transmission spectra of the PLA stack and the reference spectrum. $C_{in}$ and $C_{out}$ are the coupling and outcoupling efficiencies of the probing THz beam at the two PLA stack/air interfaces, which are presumed the same for stacks of different lengths. The RI and absorption coefficient of the PLA plates are then deducted from Fig. S1d and S1e via standard interpretation of the cutback data to be the following:

$$n_{real}(\omega) = 1.636 - 0.0331 \cdot \omega \text{ [THz]} \quad \text{(S4)}$$

$$\alpha(\omega) \text{ [cm}^{-1}\text{]} = 0.16482 + 19.695 \cdot (\omega[\text{THz}])^2 \quad \text{(S5)}$$

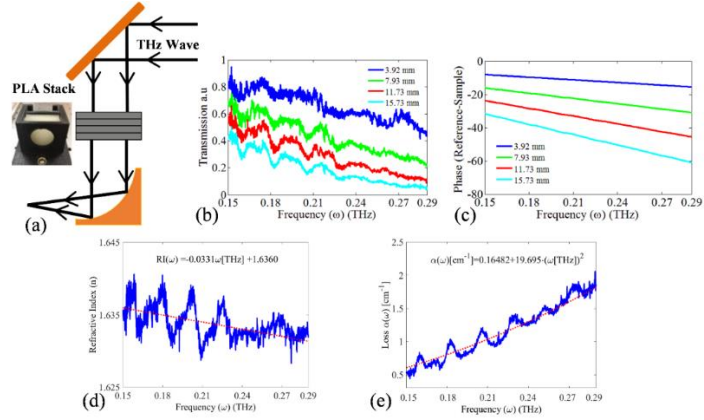

Fig. S1. (a) Schematic of the experimental setup for measuring the RI and absorption loss of PLA material used in FDM. (b) Normalized transmission spectra of the PLA stack consisting of different numbers of the PLA plates. (c) Unwrapped phase of the transmitted THz wave (relative to the reference spectrum) of the PLA stack consisting of different numbers of the PLA plates. (d) Measured and fitted RI of the PLA in 0.15 – 0.29 THz frequency range. (e) Measured and fitted absorption loss of the PLA in 0.15 – 0.29 THz frequency range.

## 8. The RI of mineral and cinnamon essential oil

The RI of the mineral and cinnamon essential oil used as liquid analytes for sensing are measured by employing cutback method and a THz-TDS system. As shown in Fig. S2a, a THz-transparent plastic container is first placed in the path of a collimated THz beam. Then, several consecutive measurements of the transmitted complex field as a function of frequency are performed by injecting oil analyte into the container successively. Based on the Eq. (S1) and (S2), the analyte RI can then be deduced from the phases of the normalized transmission spectra by using transmission through an empty container as a reference (see Fig. S2b and S2c). Under such condition, the RIs of the mineral and cinnamon essential oil are computed as 1.455 and 1.555 respectively (see Fig. S2d and S2e).

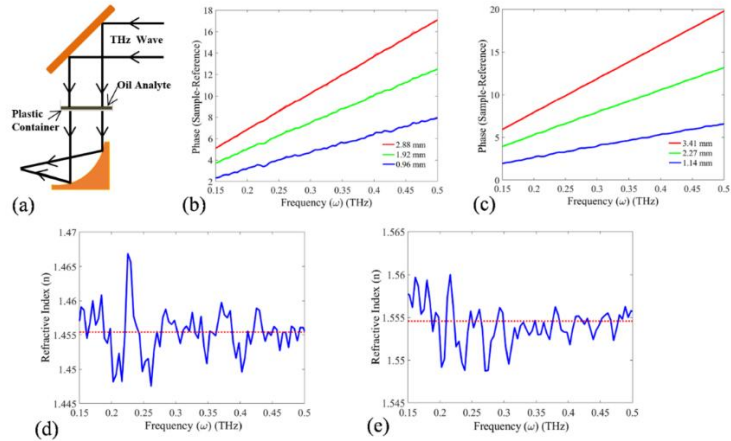

Fig. S2. (a) Schematic of the experimental setup for measuring the RI of mineral and cinnamon essential oil. (b) Unwrapped phase of the transmitted THz wave (relative to the reference spectrum) of the mineral oil with different thicknesses. (c) Unwrapped phase of the transmitted THz wave (relative to the reference spectrum) of the cinnamon essential oil with different thicknesses. (d) Measured RI of the mineral oil in the 0.15 – 0.5 THz frequency range. (e) Measured RI of the cinnamon essential oil in the 0.15 – 0.5 THz frequency range.